\documentclass[useamsfonts]{pasj01}
\usepackage{graphicx}
\usepackage{natbib}
\Received{$\langle$reception date$\rangle$}
\Accepted{$\langle$acception date$\rangle$}
\Published{$\langle$publication date$\rangle$}

\begin{document}

\title{Millimeter-sized grains in the protostellar envelopes: where do they come from?}
\author{Yi Hang Valerie \textsc{Wong},\altaffilmark{1,2,3}
Hiroyuki \textsc{Hirashita},\altaffilmark{1} and
Zhi-Yun \textsc{Li}\altaffilmark{4}}%

\altaffiltext{1}{Institute of Astronomy and Astrophysics, Academia Sinica, PO Box 23-141, Taipei 10617, Taiwan}
\altaffiltext{2}{Department of Atmospheric Sciences, National Central University, Jhongda Rd.\ 300, Jhongli 32001, Taiwan}
\altaffiltext{3}{Institute of Astronomy and Department of Physics, National Tsing Hua University, 101 Section 2 Kuang Fu Road, Hsinchu 30013, Taiwan}
\altaffiltext{4}{Astronomy Department, University of Virginia, Charlottesville, VA 22904, USA}

\email{valeriew510@gmail.com}


\maketitle

\begin{abstract}
Grain growth during star formation affects the physical and chemical processes in the evolution of star-forming clouds. We investigate the origin of the millimeter (mm)-sized grains recently observed in Class I protostellar envelopes. We use the coagulation model developed in our previous paper and find that a hydrogen number density of as high as $10^{10}~{\rm cm^{-3}}$, instead of the typical density $10^5~{\rm cm^{-3}}$, is necessary for the formation of mm-sized grains. Thus, we test a hypothesis that such large grains are transported to the envelope from the inner, denser parts, finding that gas drag by outflow efficiently ``launches'' the large grains as long as the central object has not grown to $\gtrsim 0.1$ M$_{\odot}$. By investigating the shattering effect on the mm-sized grains, we ensure that the large grains are not significantly fragmented after being injected in the envelope. We conclude that the mm-sized grains observed in the protostellar envelopes are not formed in the envelopes but formed in the inner parts of the star-forming regions and transported to the envelopes before a significant mass growth of the central object, and that they survive in the envelopes.
\end{abstract}

\section{Introduction}

Dust plays an important role in star formation. In the collapse of a molecular cloud by its self-gravity, thermal processes, especially the cooling of the gas, is of fundamental importance. Dust grains help the formation of molecular hydrogen (H$_2$) on their surfaces \citep[e.g.][]{gould63,cazaux04}. H$_2$ is one of the most important species for the thermal evolution in the protostellar collapse, which depends on metallicity and gas density \citep[e.g.][]{omukai00}. Moreover, at densities $\gtrsim 10^5$ cm$^{-3}$, dust plays an important role in cooling by radiating away the energy obtained from gas--grain collisions \citep[e.g.][]{hollenbach79,omukai00}. Both H$_2$ formation and dust cooling are governed by the total dust surface area; therefore, dust grain size distribution is crucial in determining the rates of these processes.

\citet{omukai2009} considered the effects of coagulation on the thermal
evolution of star-forming clouds. Although the dense environment enhances
the grain--grain sticking (i.e.\ coagulation) rate, coagulation affects
H$_2$ formation and dust cooling only after these
processes are not effective any more. Hence, they concluded that
the modification of grain size distribution by coagulation is not important in the thermal
evolution of collapsing clouds. According to their results, grains as
large as 1 $\micron$ form only at densities $\gtrsim 10^{13}$ cm$^{-3}$
in the solar-metallicity case, assuming the Brownian grain motion.

Some observations have shown the existence of grains larger
than $\sim 1~\micron$ in molecular cloud cores and protostellar
envelopes. Micrometer ($\micron$)-sized dust grains are shown to exist in
the dense ($\sim 10^5$ cm$^{-3}$) regions of molecular cores by the scattering-dominated features at wavelengths of a few $\micron$ -- this scattering phenomena is called coreshine \citep{pagani10,steinacker10,lefevre14}. \citet{hirashita2013}, based on their coagulation model, concluded that coagulation should last longer than a free-fall time even if they considered enhanced grain motions by turbulence \citep[see also][]{ormel2009}. This means that the molecular cloud cores are sustained against the self-gravity, creating a favorable condition for grain growth.

Grains grow effectively in circumstellar disks at a later stage of star formation \citep{natta07,li14}.
The compact emission components near the protostars ($<$a few hundreds of AU) show a small millimeter (mm) opacity spectral index ($\beta\ltsim 1$) \citep{jorgensen07}, while the dust in the interstellar medium (ISM) has $\beta\simeq 2$
\citep{draine84}. The small value of $\beta$
indicates either growth of dust to mm sizes or large optical depth.
Protoplanetary disks of Class II young stellar objects show $\beta\simeq 0$--1, which may support grain growth to mm sizes in protoplanetary disks
\citep{ricci10a}.

Some studies have shown that the mm opacity spectral index is small
($\beta <1$) even in the envelopes of Class 0 protostars 
\citep{kwon09,chiang12}.
This implies that mm-sized grains exist not only in the vicinity of protostars
but also on larger scales of protostellar envelopes.
However, since Class 0 objects are highly embedded, it is difficult to
completely separate the optically thick emission, which also decreases the apparent $\beta$.
Recently, \citet{miotello2014} have found a small $\beta (<1)$ emission at
mm wavelengths in the envelopes of two Class I objects, in which we can more easily separate the optically thin envelope component. Their results can be taken as evidence for the existence of mm-sized grains in protostellar envelopes. \citet{schnee2014} have also analyzed the mm continuum in the Orion Molecular Cloud and found a flat wavelength dependence of dust emission at mm wavelengths, which may also indicate the presence of mm-sized grains. In Fig.\ 4 of \citet{testi14}, $\beta <1$ at $\lambda \sim$ 1 mm results from a maximum grain size of $a_{\rm max} \gtrsim$ 0.3 mm for compact bare grains.
It indicates that grain growth up to $\sim 300~\micron$ is the minimal requirement to
explain the opacity change at mm wavelengths.
Some studies show that growth up to 1 mm is necessary to explain the opacity
index between $\lambda =1$ and 3 mm \citep{ricci10}. If we consider highly porous grains, the grain size should be larger than 1 mm; \citet{kataoka14} have shown that the absorption opacity of porous grains are the same if the radius times the filling factor are the same. Since the
conclusion depends on the assumptions on grain properties and radiative
transfer effects, we investigate grain growth up to $a=300~\micron$, as a minimum requirement for the phenomenon, and expand our discussions
to larger grain radii such as $a=1$--5 mm.
We refer grains with a size range of $a\sim 300~\micron$ to $5~{\rm mm}$ as mm-sized grains in this paper.

The existence of mm-sized grains at the envelope density ($\sim 10^5~{\rm cm}^{-3}$) enhances the importance of coagulation, more than expected by \citet{omukai2009} and \citet{ormel2009}. As mentioned above, the grain surface area modified by coagulation could affect the grain-surface chemistry and dust cooling in star-forming clouds. Therefore, clarifying the origin of mm-sized grains is of fundamental importance in the understanding of the star formation process.
In this paper, we seek to develop a physical scenario that explains the existence of mm-sized grains in the protostellar envelopes.

To this aim, we first apply our previous coagulation model
\citep{hirashita2013} to the formation of mm-sized grains in Section \ref{sec:coag}. We find that it is unlikely for the grains to grow to mm size in situ in the (relatively low density) envelope; additional physical processes are needed to explain the existence of mm-sized grains in protostellar envelopes. For an additional physical process, we consider the transportation of mm-sized grains from dense regions to the protostellar envelopes in
Section \ref{sec:trans}. The survival of such large grains in the envelopes is further examined in Section \ref{sec:survival}. Finally, Section \ref{sec:summary} summarizes the scenarios and the main conclusions.

\section{Grain Growth}\label{sec:coag}

In order to examine the possibility of grain growth in the Class~I envelopes,
we use the coagulation model in \citet{hirashita2013}.
We adopt a simple model by \citeauthor{hirashita2013} rather than a more comprehensive one in \citet{ormel2009} because it is much faster to compute and more suitable for the large parameter survey needed to determine the minimum density required for grain growth to mm-size.
Since we use the same models for the grain velocity dispersion
as adopted in \citet{ormel2009},
we expect that we will obtain similar results to theirs.
In the model by \citeauthor{hirashita2013}, the coagulation cross section of a pair of grains with radii $a_1$ and $a_2$ is written as
$\sigma_{12}=S\pi(a_1+a_2)^2$,
where $S\ge1$ represents the increase of the cross section by non-compact aggregates. The collision rate between the grains is estimated by the radius ($a$)-dependent grain velocities,
which are determined by the
velocity dispersion of the smallest eddies whose drag force works
efficiently within the turn-over time:
\begin{eqnarray}
v(a) & = & 1.1\times 10^3\,\left(
\frac{T_\mathrm{gas}}{10~\mathrm{K}}\right)^{1/4}
\left(\frac{a}{0.1~\micron}\right)^{1/2}\nonumber\\
& \times & \left(
\frac{n_\mathrm{H}}{10^5~\mathrm{cm}^{-3}}\right)^{-1/4}
\left(\frac{\rho_\mathrm{gr}}{3.3~\mathrm{g~cm}^{-3}}\right)^{1/2}~
\mathrm{cm~s}^{-1},\label{eq:v_turb}
\end{eqnarray}
where $T_\mathrm{gas}$ is the gas temperature,
$n_\mathrm{H}$ is the hydrogen number density,
and $\rho_\mathrm{gr}$ is the grain material density. We assume
$T_\mathrm{gas}$ to be 10 K and $\rho_\mathrm{gr}$ to be
3.3 g cm$^{-3}$ (based on silicate; \citealt{hirashita2013})
in this paper. We assume that the thickness of possible ice coating
on the surface is thin and it does not affect $\rho_\mathrm{gr}$.
Thermal velocities are small enough to be neglected.

We adopt their ``maximal coagulation model'' ($S=5$; this value is
based on \citealt{ormel2009}), which maximizes the possibility of
grain growth, so grains are coated by ice and are sticky as in
\citet{ormel2009}.
Although some authors have shown that more fluffy aggregates (i.e., larger $S$) than those in \citet{ormel2009} may form in
protoplanetary disks \citep{suyama12,okuzumi12,kataoka13,krijt15}, it is not clear whether such
extremely fluffy aggregates form before the disk formation. In particular, we consider grain collisions with velocities up to $\sim 100~{\rm m~s^{-1}}$; collisions with such velocities would lead to compaction \citep{ormel2009}.
At such a velocity, however, the effect of shattering could become a problem, making the formation of large grains even more difficult \citep{blum93,wada09,guttler10,gundlach15}.
Since the threshold velocities for shattering
such as 80 m s$^{-1}$\citep{wada13} are comparable
to the typical velocities achieved in our condition,
more detailed treatment of grain velocities by considering, for
example, grain velocity dispersions would be necessary. Thus,
we simply neglect shattering and concentrate on surveying
the most optimistic case for the formation of mm-sized grains
in this paper. Note that $a$ is defined in such a way that
$(4/3)a^3\rho_\mathrm{gr}$ is the grain mass, and that the
coagulation equations are solved for the grain mass rather than
the grain radius in our formulation.
Since the grain radius $a=300~\micron$
appropriate for ``mm-sized grains'' is defined for compact grains
(Introduction), we eventually need compaction
for consistency, or we need much larger grains to change
$\beta$ at mm wavelengths if the grains are fluffy.
Compaction may occur in the final stage of coagulation
when the grain velocities are high \citep{ormel2009},
or in the transportation
considered in Sections \ref{sec:trans} and
\ref{sec:survival}.
Since our purpose here is to estimate
the most optimistic condition of the density for
the formation of mm-sized
grains, we simply assume that such compaction
occurs before they are observed in the envelope.

The initial grain-size distribution is set to be proportional to $a^{-3.5}$, with $a_{\rm min}=0.001~\mu{\rm m}$ and $a_{\rm max}=0.25~\mu{\rm m}$, which is typical in diffuse ISM \citep*{mathis1977}:
$n(a)=\mathcal{C} a^{-3.5}~(a_{\rm min}\le a \le a_{\rm max})$,
where $\mathcal{C}$ is the normalizing constant determined from
the mass density of the grains in the ISM:
\begin{equation} \label{eq3}
\mathcal{D} \mu m_{\rm H} n_{\rm H} = \int_{a_{\rm min}}^{a_{\rm max}} \frac{4}{3} \pi a^3\rho_\mathrm{gr}\mathcal{C} a^{-3.5}\mathrm{d}a,
\end{equation}
where $m_{\rm H}$ is the hydrogen atom mass, $\mu$ is the atomic weight per hydrogen (assumed to be 1.4), and $\mathcal{D}$ (0.01: \citealt{ormel2009}) is the dust-to-gas mass ratio.
The density of the envelope is assumed to be $n_{\rm H}=10^5~{\rm cm^{-3}}$ according to \citet{miotello2014}.
We fix $n_\mathrm{H}$ and the time is always compared with the local free-fall time
($t_\mathrm{ff}=1.38\times 10^5(n_\mathrm{H}/10^5~\mathrm{cm}^{-3})^{-1/2}$ yr). Grains are assumed to be coupled with turbulence whose
eddy size is smaller than the Jeans length (so that the assumption of
constant $n_\mathrm{H}$ is reasonable) \citep{ormel2009}.

\begin{figure}
 \includegraphics[width=0.45\textwidth]{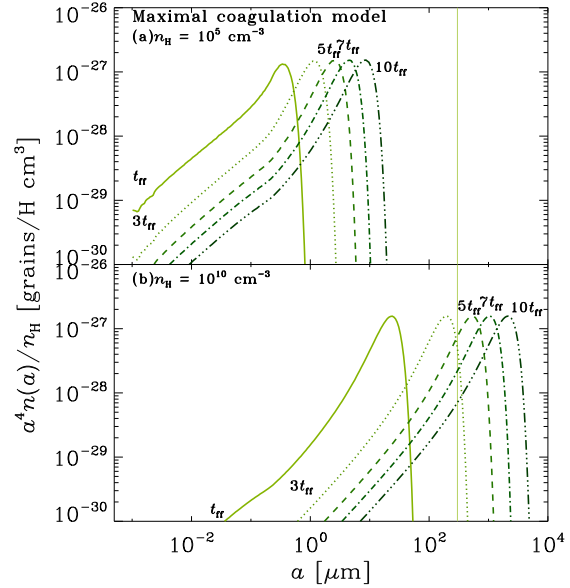}
 \caption{Evolution of grain-size distribution by coagulation. The grain size distribution is multiplied by $a^4$ to show the mass distribution per logarithmic grain radius, and is normalized to the hydrogen number density $n_\mathrm{H}$. The solid, dotted, dashed, dot-dashed and dot-dot-dot-dashed lines show the grain-size distributions at $t/t_\mathrm{ff}=$1, 3, 5, 7, and 10, respectively. The thin solid vertical line indicates the position of $300~\micron$. In panels (a) and (b), $n_{\rm H}=10^5$ (the typical density of the envelopes) and $10^{10}~{\rm cm^{-3}}$ are adopted, respectively.}
 \label{fig:coag}
\end{figure}

The evolution of grain size distribution by coagulation is shown in Fig.\ \ref{fig:coag}. As shown in Fig.\ 1a, it takes \textit{much} more than 10$t_{\rm ff}$ for the grains to grow to $>300~\mu{\rm m}$ even in the maximum coagulation model. Such a long lifetime would require an extremely long-standing sustaining mechanism of molecular clouds against gravitational collapse, and would be physically difficult. Thus, we consider another possibility for the formation of mm-sized grain.

It is easier for mm-sized grains to form in a denser region. We also show the result for $n_\mathrm{H}=10^{10}~{\rm cm^{-3}}$ in Fig.\ \ref{fig:coag}b, trying to see if mm-sized grains can form in the inner, denser regions of the molecular cloud cores. This time, grains can grow up to $\sim300~\mu{\rm m}$ in a few $t_{\rm ff}$, which is reasonably short.
Although we neglect shattering, we point out that such a
dense region is favorable for growth because grain velocities
are $\sim 30$ m s$^{-2}$ (equation \ref{eq:v_turb}), which is below the shattering
threshold \citep{wada13}.

Thus, we conclude that coagulation is not capable of forming
mm-sized grains in the envelope. In order to form
such large grains, a high density of $n_{\rm H} \gtrsim 10^{10}~{\rm cm^{-3}}$ is necessary.
This means that the existence of mm-sized grains in the protostellar envelopes
demands mechanisms other than the in situ formation.
Based on that conclusion, we hypothesize that large grains are formed in denser regions near the central object, and are subsequently transported to the envelopes.
We assume that these denser regions hosting coagulation have a spherical geometry for simplicity.
We could in principle also consider other geometries, such as a protoplanetary disk, which should
be the site of grain growth leading to the planet formation
\citep[e.g.,][]{testi14}. The basic idea remains qualitatively
the same, as long as the formed mm-sized grains are transported relatively
early, before the mass of the compact object (star + disk) becomes too large, in order to minimize
the effect of gravity, which counteracts the outward transportation.
The possibility of transportation is investigated in the next section.

\section{Transportation of Grains}\label{sec:trans}

Jets and outflows are ejected from the central circumstellar region into the envelope in the Class 0 and Class I stages (e.g. \citealt{bontemps1996}; \citealt{machida2013}). In particular, outflows, because of their large opening angles, potentially disperse the large grains in the envelope. To test this possibility, we consider the motion of mm-sized grains in the outflows.


In Section 2, we concluded that mm-sized grains only form in a region as dense as $n_{\rm H} \gtrsim 10^{10}~{\rm cm^{-3}}$. Below we consider the motion equation of a grain, whose motion is caused by the friction (drag) of the outflow and the gravitational force toward the center.
For simplicity, we assume that the outflow has a constant solid angle of $\Omega$, with a constant velocity $v_{\rm gas}$.
However, as shown below, the grain motion is determined within the central tens of AU, so that the detailed structure of outflow on scales $\gtrsim 100$ AU does not affect our conclusion.
We neglect forces originating from
non-spherical motions such as centrifugal force and
drag force from a gas rotating around the central star. The former tends to make the
outward transportation of the grains easier, while the latter causes a falling motion toward the central
star \citep{adachi76}. For the treatment of these forces, we need
to solve the growths of the disk and the central star. The
physical processes in the growths of these components depend on various
factors such as magnetic field strength, and are not fully understood yet
\citep{li14}. Thus, we leave a simultaneous treatment of the stellar
mass growth processes with grain growth to the future work.

In some numerical simulations \citep[e.g.,][]{machida2013}, we see that the mass loss rate $\dot M$ is around $0.1~{\rm M_{\odot}}/10^5~{\rm yr} \simeq 10^{-6}~{\rm M_{\odot}~yr^{-1}}$. We adopt this value for a fiducial case but examine different values for $\dot{M}$. With the mass conservation law,
\begin{equation} \label{eq4}
\dot M=\Omega r^2 \rho v_{\rm gas},
\end{equation}
where $\rho$ is the mass density of molecular gas in the outflow and $r$ is the radius from the central protostar, we get the number density of hydrogen nuclei as a function of $r$ as
\begin{eqnarray} \label{eq5}
n_{\rm H}(r) & = &\frac{\rho}{\mu m_{\rm H}}=\frac{\dot M}{\Omega v_{\rm gas} \mu m_{\rm H}}\frac{1}{r^2}\nonumber \\
       & = & 10^{10}~{\rm cm^{-3}}\left(\frac{\Omega}{4\pi /10}\right)^{-1}\left(\frac{\dot M}{10^{-6}~{\rm~M_{\odot}~yr^{-1}}}\right)\nonumber\\
       & & \times\left(\frac{v_{\rm gas}}{1~{\rm km~s^{-1}}}\right)^{-1} \left(\frac{r}{9.78~{\rm AU}}\right)^{-2}.
\end{eqnarray}

The initial radius $r_{\rm 0}$ is set at
$n_{\rm H}=10^{10}~{\rm cm^{-3}}$ ($9.78~{\rm AU}$ for $\dot M=10^{-6}~{\rm M_{\odot}~yr^{-1}}$, $v_{\rm gas}=1~{\rm km~s^{-1}}$) and $\Omega=4\pi/10$, i.e., the formation site of the mm-sized grain.
Our reference solid angle of $4\pi/10$ corresponds to an outflow opening angle of about 50 degrees, which is roughly in agreement with the observed opening angles of the winds in Class 0--I sources by \citeauthor{arce06} (\citeyear{arce06}, see their Fig.\ 5).
The drag force on the dust grain is estimated for supersonic motion as \citep{adachi76,mckee1987}
\begin{equation} \label{eq6}
F_{\rm drag}=\mu m_{\rm H}n_{\rm H}\pi a^2|v_{\rm r}|v_{\rm r},
\end{equation}
where $a$ is the radius of the dust grain and $v_{\rm r}=v_{\rm gas}-v_{\rm gr}$ is the relative velocity of the gas to that of the grain.
Noting that the grain mass is expressed as
$\frac{4}{3}\pi a^{3}\rho_\mathrm{gr}$, and including the force of gravity by the central object, we obtain the motion equation of the grain (see also equation 34 in \citealt{jones1996}):
\begin{equation} \label{eq7}
\frac{\mathrm{d}v_{\rm gr}}{\mathrm{d}t}=\frac{3\mu m_{\rm H} n_{\rm H}}{4a\rho_\mathrm{gr}}(v_{\rm gr}-v_{\rm gas})^2 - \frac{G M(<r)}{r^2},
\end{equation}
where $G$ is the gravitational constant and $M(<r)$ is the mass inside radius $r$.

First, we consider the condition that the grain obtains
an outward motion. Equating the friction term and the gravity term
in equation (\ref{eq7}) at the initial position ($v_\mathrm{gr}=0$),
and further using equation (\ref{eq5}),
we obtain the following critical mass, $M_\mathrm{cr}$:
\begin{eqnarray} \label{eq7a}
M_\mathrm{cr} & = &
\frac{3v_\mathrm{gas}\dot{M}}{4Gas\Omega}\nonumber\\
& \simeq & 2.86\times10^{-1}\,{\rm~M_{\odot}}
\left(\frac{\dot M}{10^{-6}\,{\rm~M_{\odot}\,yr^{-1}}}\right)
\left(\frac{v_\mathrm{gas}}{1\,\mathrm{~km\,s}^{-1}}\right)
\nonumber\\
& \times & \left(\frac{a}{300~\micron}\right)^{-1}
\left(\frac{\rho_\mathrm{gr}}{3.3~\mathrm{g~cm}^{-3}}\right)^{-1}
\left(\frac{\Omega}{4\pi /10}\right)^{-1};
\end{eqnarray}
that is, for $M<M_\mathrm{cr}$, the grain is transported outwards.
Thus, before the mass of the central object
(the central object effectively includes the inner disk as well as the central
protostar) grows up to 0.1--$1~{\rm M_{\odot}}$,
the grains with $a\sim 0.3$--1 mm can be ``launched'' into the envelope by the drag force of the gas outflow.
The critical mass does not depend on the radius; this is because
both drag and gravity scale with $r^{-2}$. Therefore, the condition
of grain transportation does not depend on the footpoint of the outflow
and is determined by the mass loss rate ($\dot{M}$),
velocity ($v_\mathrm{gas}$) and opening solid angle ($\Omega$) of the outflow.

\begin{figure}
 \includegraphics[width=0.45\textwidth]{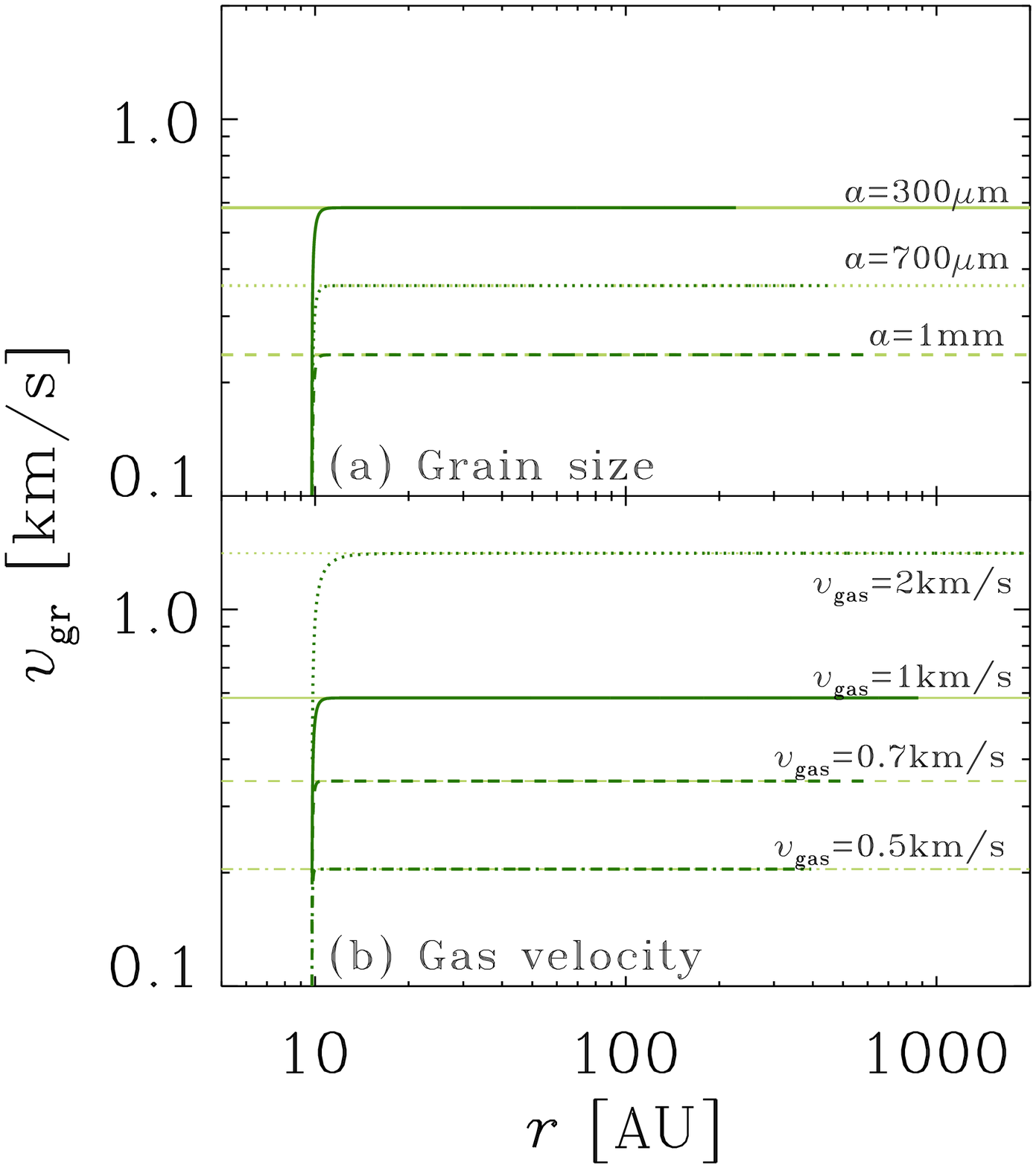}
 \includegraphics[width=0.45\textwidth]{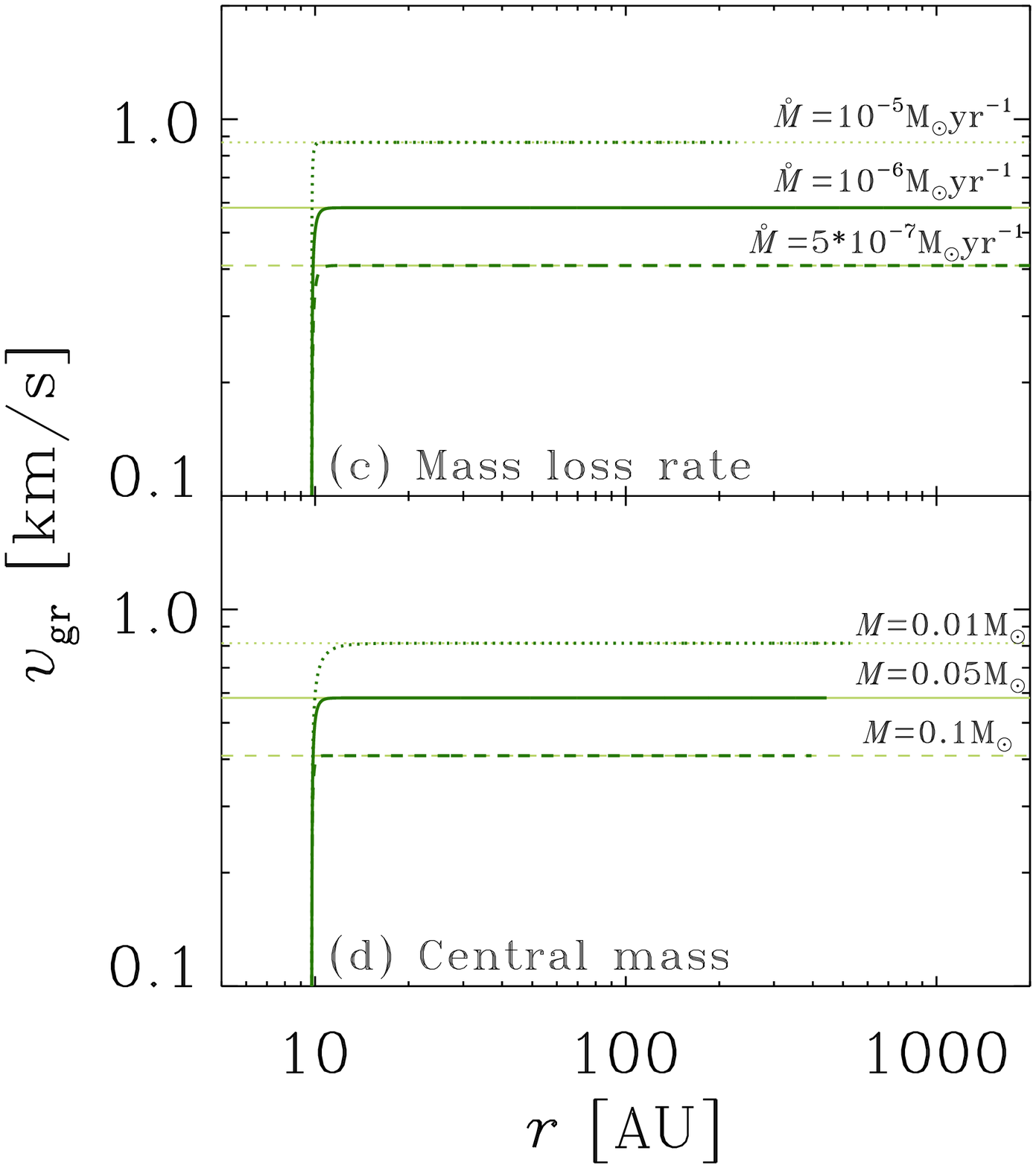}
 \caption{Grain velocity $v_{\rm gr}$ versus the radial distance
 $r$ from the central star. We choose $a=300~\mu{\rm m}$, $v_{\rm gas}=1~{\rm km~s^{-1}}$,
$\dot M=10^{-6}~{\rm M_{\odot}~yr^{-1}}$ and $M=0.05~{\rm M_{\odot}}$ for the fiducial values and change one of those parameters in each panel.
The horizontal thin lines show the analytically estimated terminal velocities
and their line species follow the legends below.
(a) The dotted, solid, and dashed lines show the relation between $v_{\rm gr}$ and $r$ for grain radius $a=300~\mu{\rm m}$, $700~\mu{\rm m}$ and $1~{\rm mm}$, respectively. (b) The dot-dashed, dashed, solid, and dotted lines show the relation for gas velocity $v_{\rm gas}=0.5$, $0.7$, $1.0$ and $2.0~{\rm km~s^{-1}}$, respectively. (c) The dotted, solid, and dashed lines show the relation for the mass loss rate $\dot{M}=10^{-5}$, $10^{-6}$, and $5\times 10^{-7}$ M$_{\odot}$ yr$^{-1}$, respectively. (d) The dotted, solid, and dashed lines show the relation for the mass of the central object ${M}=0.01$, $0.05$, and $0.1$ M$_{\odot}$, respectively.}
\label{fig:transport}
\end{figure}

To investigate the efficiency of grain acceleration, we solve equation (\ref{eq7}) focusing on the case of $M<M_\mathrm{cr}$.
For simplicity, we assume that the mass is dominated by the central
region (i.e., inside the initial radius of the grain) and we fix the central mass.
For illustration, we adopt $a=300~\mu{\rm m}$, $v_{\rm gas}=1~{\rm km~s^{-1}}$, $\dot M=10^{-6}~{\rm M_{\odot}~yr^{-1}}$ and $M=0.05~{\rm M_{\odot}}$ for the fiducial case. We solve equation (\ref{eq7}), varying one parameter at a time, with the others fixed at the fiducial values. We obtain the grain velocity $v_{\rm gr}$ as a function of radius $r$ as shown in Fig.\ \ref{fig:transport}.

Fig.\ \ref{fig:transport} shows that acceleration occurs locally
around the initial radius, which justifies the assumption of fixing $M$.
Moreover, since both the drag and the gravity terms have the same dependence on
$r$, the grains converge to a terminal velocity determined by
equating the two terms (shown by the horizontal lines in Fig.\ \ref{fig:transport}).
As shown in Fig.\ \ref{fig:transport}a, with increasing grain radius $a$, the terminal velocity of the grain decreases. This is because with a larger grain size, the surface-to-volume (mass) ratio of the grain is smaller, so that the acceleration by drag is less efficient. However, even mm-sized grains are coupled with the gas strongly enough to attain a velocity of $\gtrsim 0.2~{\rm km~s^{-1}}$ around the initial radius.
The time-scale of the grain transport is estimated as
$\sim 1000~{\rm AU}~/~0.5~{\rm km~s^{-1}} \sim 9480~{\rm yr}$, which is shorter than the free-fall time at the typical density ($n_{\rm H} \sim 10^5~{\rm cm^{-3}}$) of the envelope ($\sim 10^{5}~{\rm yr}$) and can be shorter than the time-scale of
the mass growth of the central star ($\lesssim 10^5$ yr;
\citealt{inutsuka2010}).

In Fig.\ \ref{fig:transport}b, we observe that the grain is
efficiently accelerated outwards
in all the relevant range of $v_\mathrm{gas}$. There are two competing
effects of varying $v_\mathrm{gas}$. If $v_\mathrm{gas}$ is high,
the effect of drag relative to that of gravity is enhanced. At the same time,
a greater gas velocity also leads to a lower gas density (equation \ref{eq5}),
thus a weaker drag.
Yet in all cases, the grains are efficiently accelerated outwards.

In Fig.\ \ref{fig:transport}c, with increasing mass loss rates, grains are more strongly coupled with the gas. This is because a higher mass loss rate with a fixed gas velocity results in a higher density (equation \ref{eq5}), and thus a stronger drag. For $a=300~\micron$,
even the case of mass loss rate as small as
$5\times10^{-7}~{\rm M_{\odot}~yr^{-1}}$ achieves a terminal velocity of $0.41~{\rm km~s^{-1}}$, and the acceleration occurs locally at around the initial radius.

Finally, as shown in Fig.\ \ref{fig:transport}d, an increase of the central mass results in lower coupling of grains with gas. With larger central mass, the drag term is less dominated. Nonetheless, even a central mass of as large as $0.1~{\rm M_{\odot}}$ gives rise to outward transportation of grains, resulting in efficient acceleration to a terminal velocity of $0.41~{\rm km~s^{-1}}$ around the initial radius.


\begin{figure}
 \includegraphics[width=0.45\textwidth]{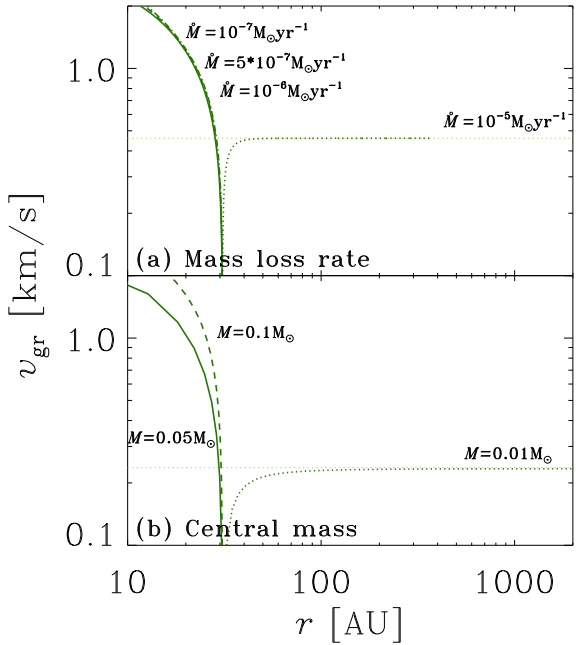}
 \caption{Same as Fig.\ \ref{fig:transport} but for $a=5~{\rm mm}$.
 (a) The dotted, solid, dashed, and dot-dashed lines show the relation for the mass loss rate $\dot{M}=10^{-5}$, $10^{-6}$, $5\times 10^{-7}$, and $10^{-7}$ M$_{\odot}$ yr$^{-1}$, respectively.
 The grain obtains an outward motion only in the case of $\dot{M}=10^{-5}$ M$_\odot$ yr$^{-1}$ and falls inward in the other cases.
 (b) The dotted, solid, and dashed lines show the relation for the mass of the central object ${M}=0.01$, $0.05$, and $0.1$ M$_{\odot}$, respectively.
 Outward motion is only possible in the case of $M=0.01$ M$_\odot$.
 }\label{fig:5mm}
\end{figure}

We further explore the case for $a=5$ mm. Setting the grain radius as $a=5~{\rm mm}$, we change the other parameters as we did in Fig.\ \ref{fig:transport}. Results are shown in Fig.\ \ref{fig:5mm}. For grains as large as $5~{\rm mm}$, gas velocities ranging from $v_{\rm gas}=0.5$--$2~{\rm km~s^{-1}}$ are no longer able to transport the 5-mm grains if we adopt the fiducial values for the other parameters. Thus, we do not show the plot for various $v_\mathrm{gas}$. Considering the effect of mass loss rate, as shown in Fig.\ 3a, transportation of grains with $a=5$ mm is only possible if $\dot M$ is as large as $10^{-5}~{\rm M_{\odot}~yr^{-1}}$. Furthermore, from equation (\ref{eq7a}), since the critical mass of grains with $a=5$ mm becomes $M_{\rm cr} \sim 0.017~{\rm M_{\odot}}$, such large grains can only be transported outward if the central mass is smaller than the critical mass; i.e.\ for the $M=0.01~{\rm M_{\odot}}$ case in Fig.\ 3b.

In conclusion, despite the variation of grain size $a$, gas velocity $v_{\rm gas}$, mass loss rate $\dot M$ and central mass $M$, grains with $a\sim 300~\micron$--$1$ mm are successfully accelerated to a velocity similar to the outflow on a spatial scale of $\sim 10$ AU.
Moreover, the time-scale of transportation to the envelope ($\sim 1000~{\rm AU}$) from a dense ($n_{\rm H} \sim 10^{10}~{\rm cm^{-3}}$) region, where the growth of grains to mm sizes is possible, can be shorter than that of the central mass growth to $M_\mathrm{cr}\sim 0.1$--$0.3~$M$_\odot$.
Even the grains with the radius as large as 5 mm can be transported outward
if the mass loss rate of the outflow is as large as $10^{-5}$ M$_\odot$ yr$^{-1}$
or the central object is light enough ($M\lesssim 0.01$ M$_\odot$).

\section{Survival of grains}\label{sec:survival}
Now
we examine whether mm-sized grains transported from the denser part of the protostellar cores to the envelope can survive in the envelope or not.
Since the grain velocity predicted is as large as $1~{\rm km~s^{-1}}$, the grain may be shattered in collisions with other grains. To estimate whether or not the mm-sized grains survive against shattering, we consider their collisions with other grains before their motions are significantly decelerated by gas friction in the envelope.

\subsection{Formulation}
After being injected into the envelope, the mm-sized grain travels in
the envelope on the friction time-scale, $\tau_{\rm f}$, estimated by dividing the grain momentum ($\sim \frac{4}{3} \pi a_0^3 s v_{\rm gr}$, where $a_0$
is the radius of the mm-sized grain) by friction ($\sim \pi a_0^2 v_{\rm gr} \mu m_{\rm H}$):
\begin{equation} \label{eq8}
\tau_{\rm f} =\frac{4 a_{\rm 0}\rho_\mathrm{gr}}{3 v_{\rm gr} \mu m_{\rm H} n_{\rm H}}.
\end{equation}
For $v_{\rm gr}$, we adopt $\sim$1 km s$^{-1}$ based on the terminal velocities calculated in Section 3.

Now we consider the effect of shattering in a friction time. We assume that the size distribution of colliding grains, referred to as the field grains, is described by a power-law of index $p$, $Ca^{-p}$ ($C$ is the normalizing constant). The normalization factor $C$ is determined from the total dust mass in the column swept by the mm-sized grain in a friction time:
\begin{equation} \label{eq9}
\mathcal{D} \mu m_{\rm H} n_{\rm H} \pi a_{\rm 0}^2 v_\mathrm{gr} \tau_{\rm f} = \int_{\tilde{a}_{\rm min}}^{\tilde{a}_{\rm max}} \frac{4}{3} \pi a^3\rho_\mathrm{gr}C a^{-p} \mathrm{d}a,
\end{equation}
with the minimum and maximum radii of $\tilde{a}_{\rm min}$ and $\tilde{a}_{\rm max}$ respectively.
We assume that $\tilde{a}_{\rm max}=a_0$, but the choice of $\tilde{a}_{\rm max}$ does not affect the conclusion as shown in Section 4.2.

In finding the shattered fraction of the mm-sized grain, we make use of the shattering model in \citet{kobayashi2009}. The normalized impact energy $\phi$ (equation 8 of \citealt{kobayashi2009}) is defined as
\begin{equation} \label{eq11}
\phi (y) = \frac{v^2}{2Q_{\rm D}^*}\frac{y}{1+y},
\end{equation}
where $Q_{\rm D}^*$ is the threshold impact energy per unit mass of
the mm-sized grain for catastrophic disruption (here
catastrophic disruption indicates that more than half of
the grain is fragmented), and $y\equiv m/m_0=a^3/a_0^3$, with $m$ and $a$, respectively, being the mass and radius of the field grain colliding with the mm-sized grain, and $m_0$ and $a_0$, respectively, the mass and radius of the mm-sized grain.
The mass of the ejected fragment from the mm-sized grain, $m_{\rm e}$, is given by
(see equation 10 in \citealt{kobayashi2009})
\begin{equation} \label{eq12}
m_{\rm e} = \frac{\phi (y)}{1+\phi (y)}\, m_0.
\end{equation}
Using the fraction of destruction in a single collision $m_{\rm e}/m_0=\phi/(\phi+1)$, we can estimate the contribution of the field grains with radii between $a$ and $a+\mathrm{d}a$ to the shattered fraction of the mm-sized grain, $D(a)\,\mathrm{d}a$:
\begin{equation} \label{eq13}
D(a)\,\mathrm{d}a=C a^{-p} \frac{\phi (y)}{\phi (y)+1}\,\mathrm{d}a.
\end{equation}
Note that $y=a^3/a_0^3$ is a function of $a$.
By integrating $D(a)$ from $\tilde{a}_{\rm min}$ to $\tilde{a}_{\rm max}$, we get the total destroyed fraction of the mm-sized grain, $\Phi$, in the friction time as
\begin{equation} \label{eq14}
\Phi = \int_{\tilde{a}_{\rm min}}^{\tilde{a}_{\rm max}} D(a)\mathrm{d}a.
\end{equation}

We adopt $Q_{\rm D}^* =3.5 \times 10^7(a_0/1~{\rm cm})^{-0.38}$ erg g$^{-1}$ \citep{benz1999,kobayashi2009}, which is valid for the case of compact grains.
As a comparison, we also investigate the threshold impact energy appropriate for
aggregates; $Q_{\rm D}^*=4 \times 10^7(v/100~\mathrm{m~s}^{-1})(r^\prime/0.1~\micron)^{-5/6}~{\rm erg~g^{-1}}$,
where $r^\prime$ is the grain radius composing the aggregate, as in equation (10) of \citet{wada13} (we assume that $r^\prime=0.1~\micron$).
To distinguish these two cases, we adopt the notations, $Q_{\rm D}^*=Q_\mathrm{D,c}^*$
and $Q_\mathrm{D,a}^*$, for the former and latter cases respectively.
The resulting $\Phi$ is also distinguished by the notations $\Phi_\mathrm{c}$
and $\Phi_\mathrm{a}$ correspondingly.

\subsection{Results}

\begin{figure}
\includegraphics[width=0.45\textwidth]{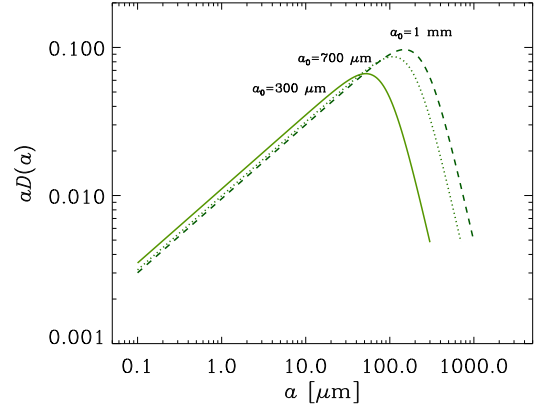}
 \caption{
Contribution to the destruction fraction of the mm-sized grain from field grains with radii between $a$ and $a+\mathrm{d}a$. To show the destruction fraction per logarithmic grain radius, we multiply $D(a)$ by $a$. The solid, dotted and dashed lines show the results for $a_0=300~\micron$, $700~\micron$ and $1~{\rm mm}$ respectively. For example, for $a_0=1$~mm, $aD(a)$ peaks at around 200 $\micron$ and the peak value is $\sim$0.1, which means that the largest contribution to the destruction of the mm-sized grain comes from the field grains with $a\sim 200~\micron$ and that around 10 per cent of the mm-sized grain is destroyed by shattering.}
\label{fig:destruction}
\end{figure}

We first consider the cases with threshold energy $Q_{\rm D}^*=Q_\mathrm{D,c}^*$.
To start with, we investigate the case for various $a_0$ with
$a_{\rm min}=0.1~\micron$,
$v_\mathrm{gr}=1$ km s$^{-1}$,
and $p=3.5$ \citep{kobayashi2009}.
In Fig.\ \ref{fig:destruction}, we plot $aD(a)$, which quantifies the contribution of the field grains per $\log a$ to the shattered fraction of the mm-sized grain, for different sizes of the mm-sized grains $a_{0}$. We observe that a peak exists for each curve, and that the peak values of $aD(a)$ are well below $1$.
For example, for $a_0=1$ mm, $aD(a)$ peaks at around $a\sim 200~\micron$ and the peak value is $\sim$0.1, which means that the largest contribution to the destruction of the mm-sized grain comes from the field grains with $a\sim 200~\micron$ and that around 10 per cent of the mm-sized grain is destroyed by shattering.
Indeed, after integration, we obtain $\Phi_\mathrm{c} =0.33$, which is roughly equal to the peak of $aD(a)$.
At small $a$, the slope is positive with a value of $(4-p)$. The reason for this is that when $\phi \ll 1$ [i.e., $a \ll (2Q_\mathrm{D,c}^*/v_\mathrm{gr}^2)^{1/3} a_0$], $D(a) \sim Ca^{-p}\phi$ (equation \ref{eq13}), the slope of $D(a)$ becomes $3-p$. The peak appears at around $\phi \sim 1$. Beyond the peak, $\phi/(\phi+1) \sim 1$, so $D(a) \sim Ca^{-p}$. The peaks are well established as long as $p<4$, which is naturally realized after multiple disruption processes \citep{kobayashi2009,kobayashi2013}. Therefore, the peak position is primarily determined by $a \sim (2Q_\mathrm{D,c}^*/v_\mathrm{gr}^2)^{1/3} a_0$ corresponding to $\phi \sim 1$.

Other parameters such as $\tilde{a}_{\rm min}$ and $p$ have minor effects on the total destroyed fraction $\Phi_\mathrm{c}$. The destroyed fraction $\Phi_\mathrm{a}$ (with $Q_{\rm D}^*=Q_\mathrm{D,a}^*$) is rather smaller than $\Phi_\mathrm{c}$ because $Q_{\rm D,c}^*<Q_\mathrm{D,a}^*$ in the parameter ranges of interest. In Table \ref{tab:Phi}, we show the dependence of $\Phi_\mathrm{c}$ and $\Phi_\mathrm{a}$ on various parameters to confirm that, unless the grain velocity is $\gtrsim~1.5~\mathrm{km~s^{-1}}$, $\Phi$ is always significantly smaller than 1. Therefore, mm-sized grains are likely to survive in the envelope.

\begin{table}

\begin{center}
\begin{tabular}{@{}lccc} \hline
Variable & Range & $\Phi_\mathrm{c}$ & $\Phi_\mathrm{a}$\\ \hline
$a_0$ & 0.3--1 mm & 0.21--0.33 & 0.083--0.086 \\
$a_\mathrm{min}$ & 0.01--1 $\micron$ & 0.23--0.20 & 0.090--0.081 \\
$v_\mathrm{gr}$ & 0.5--1.5 km s$^{-1}$ & 0.065--0.42 & 0.060--0.17\\
$p$ & 3--4 & 0.13--0.39 & 0.11--0.13 \\ \hline
\end{tabular}
\end{center}
 \caption{Shattered fractions of the mm-sized grain $\Phi_\mathrm{c}$ and $\Phi_\mathrm{a}$ with the fiducial parameter set of $(a_0,\, \tilde{a}_\mathrm{min},\, v_\mathrm{gr},\, p)=(300~\micron,\, 0.1~\micron ,\, 1~\mathrm{km~s}^{-1},\, 3.5)$. Only the parameter shown in the column of ``Variable'' is changed in the corresponding range on the right.}\label{tab:Phi}
\end{table}


\section{Summary}\label{sec:summary}

According to a recent observation by \citet{miotello2014}, millimeter (mm)-sized grains exist in the protostellar envelopes. We thus investigated the origin of such mm-sized grains using the model by \citet{hirashita2013}. We found that the typical density of the envelopes $n_{\rm H}\sim 10^5~{\rm cm^{-3}}$ is not sufficient for the formation of such large grains. Only with densities $n_{\rm H} \gtrsim 10^{10}~{\rm cm^{-3}}$ would such large grains be possible to form. This means that another mechanism is required to explain the existence of mm-sized grains in the envelopes. Therefore, we further examined the possibility that mm-sized grains formed in denser parts near the central protostar are transported to the envelope by an outflow. First, we found a critical central mass below which the drag force on the dust can win over gravity. This mass is estimated to be of order 0.1 M$_{\odot}$ for a mass loss rate of $10^{-6}~{\rm M_{\odot}~yr^{-1}}$. Before the central mass reaches this critical mass, mm-sized grains are transported to the envelope from the dense regions in $\sim10^4$ yr, which is shorter than the mass growth time-scale of the central star. Finally, we examined if the mm-sized grains injected into the envelope are able to survive after shattering in the envelope. We found that the shattered fraction is $\Phi \sim 10^{-1}$, and is robustly smaller than 1 for the reasonable ranges of relevant parameters unless the grain is injected in the envelope with high velocity $>1.5~\mathrm{km~s^{-1}}$. This means that mm-sized grains are likely to survive. We hence conclude that transportation of mm-sized grains formed in the central dense region to the envelope is a possible scenario for the appearance of mm-sized grains in the protostellar envelope.

\section*{Acknowledgments}

We thank the anonymous referee for useful comments that
improved the scientific discussions of this paper very much.
YHVW thanks Wei-Hao Wang for his support
through the the Ministry of Science and Technology
(MoST) grant 102-2119-M-001-007-MY3.
HH acknowledges the support from the MoST grant 102-2119-M-001-006-MY3.
ZYL is supported in part by NSF AST 1313083 and NASA NNX14AB38G.


\begin{thebibliography}{}
\bibitem[\protect\citeauthoryear{Adachi, Hayashi, \& Nakazawa}{Adachi et al.}{1976}]{adachi76}
Adachi I., Hayashi C., Nakazawa K., 1976, Prog.\ Theor.\ Phys., 56, 1756
\bibitem[\protect\citeauthoryear{Arce \& Sargent}{Arce \& Sargent}{2006}]{arce06}
Arce H. G., Sargent A. I., 2006, ApJ, 646, 1070
\bibitem[\protect\citeauthoryear{Benz \& Asphaug}{1999}]{benz1999} Benz W., Asphaug E., 1999, Icarus, 142, 5
\bibitem[\protect\citeauthoryear{Blum \& M\"unch}{1993}]{blum93}
Blum J., Münch M., 1993, Icarus, 106, 151
\bibitem[\protect\citeauthoryear{Bontemps et al.}{1996}]{bontemps1996} Bontemps S., Andr\'{e} P., Terebey S., Cabrit S., 1996, A\&A, 311, 858
\bibitem[\protect\citeauthoryear{Cazaux \& Tielens}{2004}]{cazaux04}
Cazaux S., Tielens A. G. G. M., 2004, ApJ, 604, 222
\bibitem[\protect\citeauthoryear{Chiang, Looney, \& Tobin}{2012}]{chiang12}
Chiang H.-F., Looney L. W., Tobin J., 2012, ApJ, 756, 168 
\bibitem[\protect\citeauthoryear{Draine \& Lee}{1984}]{draine84}
Draine B. T., Lee H. M., 1984, ApJ, 285, 89
\bibitem[\protect\citeauthoryear{Gould \& Salpeter}{1963}]{gould63}
Gould R. J., Salpeter E. E., 1963, ApJ, 138, 393
\bibitem[\protect\citeauthoryear{Gundlach \& Blum}{2015}]{gundlach15}
Gundlach B., Blum J., 2015, ApJ, 798, 34
\bibitem[\protect\citeauthoryear{G\"uttler et al.}{2010}]{guttler10}
G\"uttler C., Blum J., Zsom A., Ormel C. W., Dullemond C. P., 2010, A\&A, 513, 56
\bibitem[\protect\citeauthoryear{Hirashita \& Kobayashi}{2013}]{kobayashi2013} Hirashita H., Kobayashi H., Earth Planets Space, 65, 1083
\bibitem[\protect\citeauthoryear{Hirashita \& Li}{2013}]{hirashita2013} Hirashita H., Li Z.-Y., 2013, MNRAS, 434, L70
\bibitem[\protect\citeauthoryear{Hirashita \& Omukai}{2009}]{omukai2009} Hirashita H., Omukai K., 2009, MNRAS, 399, 1795
\bibitem[\protect\citeauthoryear{Hollenbach \& McKee}{1979}]{hollenbach79}
Hollenbach D., McKee C. F., 1979, ApJS, 41, 555
\bibitem[\protect\citeauthoryear{Inutsuka et al.}{2010}]{inutsuka2010}
Inutsuka S.-I., Machida M. N., Matsumoto T., 2010, ApJS, 718, 58
\bibitem[\protect\citeauthoryear{Jones et al.}{1996}]{jones1996} Jones A. P., Tielens A. G. G. M., Hollenbach D. J., 1996, ApJ, 496, 740
\bibitem[\protect\citeauthoryear{J{\o}rgensen et al.}{2007}]{jorgensen07}
J{\o}rgensen J. K., et al., 2007, ApJ, 659, 479
\bibitem[\protect\citeauthoryear{Kataoka et al.}{2014}]{kataoka14}
Kataoka A., Okuzumi S., Tanaka H., Nomura H., 2014, A\&A, 568, A42
\bibitem[\protect\citeauthoryear{Kataoka et al.}{2013}]{kataoka13}
Kataoka A., Tanaka H., Okuzumi S., Wada K., 2013,
A\&A, 554, A4
\bibitem[\protect\citeauthoryear{Kobayashi \& Tanaka}{2009}]{kobayashi2009} Kobayashi H., Tanaka H., 2010, Icar, 206, 735
\bibitem[\protect\citeauthoryear{Krijt et al.}{2015}]{krijt15}
Krijt S., Ormel C. W., Dominik C., Tielens A. G. G. M., 2015, A\&A, 574, A83
\bibitem[\protect\citeauthoryear{Kwon et al.}{2009}]{kwon09}
Kwon W., Looney L. W., Mundy L. G., Chiang H.-F., Kemball A., 2009,
ApJ, 696, 841
\bibitem[\protect\citeauthoryear{Lef\`{e}vre et al.}{2014}]{lefevre14}
Lef\`{e}vre C., et al., 2014, A\&A, 572, A20
\bibitem[\protect\citeauthoryear{Li et al.}{2014}]{li14}
Li Z.-Y., Banerjee R., Pudritz R. E., Jørgensen J. K., Shang H., Krasnopolsky R., Maury A., 2014,
in Protostars and Planets VI, eds.\ H. Beuther, R. S. Klessen, C. P. Dullemond, and T. Henning
(Tucson: University of Arizona Press), 173
\bibitem[\protect\citeauthoryear{Machida \& Hosokawa}{2013}]{machida2013} Machida M. N., Hosokawa T., 2013, MNRAS, 431, 1719
\bibitem[\protect\citeauthoryear{Mathis, Rumpl \& Nordsieck}{1977}]{mathis1977} Mathis J. S., Rumpl W., Nordsieck K. H., 1977, ApJ, 217, 425
\bibitem[\protect\citeauthoryear{McKee et al.}{1987}]{mckee1987} McKee C.F., Hollenbach D. J., Seab C. G., Tielens A. G. G. M., 1987, ApJ, 318, 674
\bibitem[\protect\citeauthoryear{Miotello et al.}{2014}]{miotello2014} Miotello A., Testi L., Lodato G., Ricci L., Rosotti G., Brooks K., Maury A., Natta A., 2014, A\&A, 567, A32
\bibitem[\protect\citeauthoryear{Natta et al.}{2007}]{natta07}
Natta A., Testi L., Calvet N., Henning T., Waters R., Wilner D., 2007, in Protostars and Planets V, eds.\ B. Reipurth, D. Jewitt, and K. Keil (Tucson: University of Arizona Press), 767
\bibitem[\protect\citeauthoryear{Okuzumi et al.}{2012}]{okuzumi12}
Okuzumi S., Tanaka H., Kobayashi H., Wada K., 2014,
ApJ, 752, 106
\bibitem[\protect\citeauthoryear{Omukai}{2000}]{omukai00}
Omukai K., 2000, ApJ, 534, 809
\bibitem[\protect\citeauthoryear{Ormel et al.}{2009}]{ormel2009} Ormel C. W., Paszun D., Dominik C., Tielens A. G. G. M., 2009, A\&A, 502, 845
\bibitem[\protect\citeauthoryear{Pagani et al.}{2010}]{pagani10} Pagani L., Steinacker J., Bacmann A., Stutz A., Henning T., 2010, Science, 329, 1622
\bibitem[\protect\citeauthoryear{Ricci et al.}{2010a}]{ricci10a}
Ricci L., Testi L., Natta A., Brooks K. J., 2010a, A\&A, 521, A66
\bibitem[\protect\citeauthoryear{Ricci et al.}{2010b}]{ricci10}
Ricci L., Testi L., Natta A., Neri R., Cabrit S., Herczeg G. J., 2010b, A\&A, 512, A15
\bibitem[\protect\citeauthoryear{Schnee et al.}{2014}]{schnee2014} Schnee S., Mason B., Francesco J. D., Friesen R., Li D., Sadavoy S., Stanke T., 2014, MNRAS, 444, 2303
\bibitem[\protect\citeauthoryear{Steinacker et al.}{2010}]{steinacker10}
Steinacker J., Pagani L., Bacmann A., Guieu S., 2010, A\&A, 511, A9
\bibitem[\protect\citeauthoryear{Suyama et al.}{2012}]{suyama12}
Suyama T., Wada K., Tanaka H., Okuzumi S., 2012, ApJ, 753, 115
\bibitem[\protect\citeauthoryear{Testi et al.}{2014}]{testi14}
Testi L., Birnstiel T., Ricci L., Andrews S., Blum J., Carpenter J., Dominik C., Isella A., Natta A., Williams J. P., Wilner D. J., 2014,
in Protostars and Planets VI, eds.\ H. Beuther, R. S. Klessen, C. P. Dullemond, and T. Henning
(Tucson: University of Arizona Press), 339
\bibitem[\protect\citeauthoryear{Wada et al.}{2013}]{wada13}
Wada K., Tanaka H., Okuzumi S., Kobayashi H., Suyama T., Kimura H., Yamamoto T., 2013, A\&A 559, A62
\bibitem[\protect\citeauthoryear{Wada et al.}{2009}]{wada09}
Wada K., Tanaka H., Suyama T., Kimura H., Yamamoto T., 2009, ApJ, 702, 1490
\end{thebibliography}
\end{document}